\documentclass[aps,prb,twocolumn,groupedaddress,showpacs,floatfix]{revtex4-1}
\pdfoutput=1

\usepackage[%
  final,
]{graphicx}

\usepackage[]{color}
\usepackage{bm}
\usepackage{paralist}

\usepackage[%
  pdfborder={0 0 0}, 
]{hyperref}

\definecolor{darkblue}{RGB}{0,0,90}
\hypersetup{
  colorlinks=true,
  linkcolor=darkblue,
  anchorcolor=darkblue,
  citecolor=darkblue,
  filecolor=darkblue,
  menucolor=darkblue,
  runcolor=darkblue,
  urlcolor=darkblue,
}

\usepackage{nicefrac}

\usepackage{dcolumn}
\newcolumntype{h}[1]{D{.}{.}{#1}}

\newcommand{\ket}[1]{%
\mathchoice{%
\left|#1\right\rangle}{%
|#1\rangle}{%
\left|#1\right\rangle}{%
\left|#1\right\rangle}%
}

\newcommand{\scalprod}[2]{\langle #1 \vert #2 \rangle}
\newcommand{\expect}[1]{%
\mathchoice{%
\left\langle #1 \right\rangle}{%
\langle #1 \rangle}{%
\left\langle #1 \right\rangle}{%
\left\langle #1 \right\rangle}%
}
\newcommand{\op}[1]{\hat{#1}}
\newcommand{\note}[1]{}

\newcommand{\vect}[1]{\bm{#1}}
\newcommand{\vecop}[1]{\hat{\bm{#1}}}
\newcommand{\norm}[1]{\vert #1 \vert}

\begin{document}


\title{Solitary excitations in one-dimensional spin chains}


\author{Anton W\"ollert}
\email[E-mail: ]{anton.woellert@mpi-hd.mpg.de}
\affiliation{Institut f\"ur Theoretische Physik, Georg-August-Universit\"at G\"ottingen,
37077 G\"ottingen, Germany}

\author{Andreas Honecker}
\affiliation{Institut f\"ur Theoretische Physik, Georg-August-Universit\"at G\"ottingen,
37077 G\"ottingen, Germany}
\affiliation{Fakult\"at f\"ur Mathematik und Informatik,
             Georg-August-Universit\"at G\"ottingen,
             37073 G\"ottingen, Germany}


\date{February 21, 2012; revised May 8, 2012} 

\begin{abstract}
We study the real-time evolution of solitary excitations in 1-d quantum 
spin chains using exact diagonalization (ED) and the density-matrix 
renormalization group (DMRG). The underlying question of this work is the 
correspondence between classical solitons and solitons in quantum 
mechanics. While classical solitons as eigensolutions of non-linear wave 
equations are localized and have a sharp momentum, this is not possible in 
the corresponding quantum case due to the linearity of the Schr\"odinger 
equation or seen in a more pictorial way, because of the uncertainty 
relation. For the case of the XXZ model it is shown that the real-time 
evolution of quantum wave packets accompanied by spreading is in 
qualitative accordance with the one predicted by classical solitons.
\end{abstract}

\pacs{75.10.Pq, 75.40.Mg, 47.35.Fg} %

\maketitle

\section{Introduction}
Solitons, first mentioned by John Scott Russel in 1844, are outstanding objects
in the field of nonlinear physics.\cite{scott1973soliton} Their description as
solution of nonlinear wave equations needs to take into account the full
nonlinearity of the problem. Based on the numerical findings by Kruskal and
Zabusky,\cite{zabusky1965interaction} the inverse scattering transform
(IST)\cite{gardner1967method} was the first major framework to systematically
research the solutions and spectra of integrable%
\footnote{The term ``integrable'' is well defined for classical mechanics. As
for quantum mechanics, we refer to this term in the sense of ``soluble by the
quantum inverse scattering method'' (see Ref.~\onlinecite{korepin1997quantum}).}
classical nonlinear wave equations like the sine-Gordon equation or the
nonlinear Schr\"odinger equation. These soliton solutions are usually
characterized by a constant shape and velocity which is due to cancellation of
dispersion and nonlinearity.

The extension of the term ``soliton'' to the quantum regime is not
straightforward. On the one hand there are technical problems of quantizing a
classical nonlinear wave equation to a quantum field theory. For classical
models amenable to the IST a direct canonical quantization is
possible\cite{zakharov1971korteweg} because the IST can be seen as a nonlinear
canonical mapping to action angle variables which can be directly quantized.
Quasiclassical quantization\cite{[{}][{ [Sov. J. Low. Temp. Phys. \textbf{3}, 440 (1977)].}]kosevich1977nonlinear} has also been used for
identifying classical and quantum system. But still an obvious problem seems to
exists in the quantum case. This is the interpretation of a quantum soliton.
Because a quantum soliton should not have a constant shape and velocity (due to
the uncertainty relation) as is the case for a classical soliton. Another point
of view on this problem is that classical solitons are eigensolutions of
nonlinear wave equations. These eigensolutions are localized by means of some
observable like density or magnetization in space even if the system is
translationally invariant. In the quantum case this is not possible,%
\footnote{Localized eigenstates for translationally invariant quantum models
might exist if the model has a flat band (as for quite a few frustrated
models\cite{[{}][{ [Fizika Nizkikh Temperatur \textbf{33}, 982 (2007)].}]derzhko:745,PhysRevB.82.184502}), but for the quantized versions of the
nonlinear wave equations mentioned in this work, this is not the case.
Furthermore these types of localized eigenstates cannot move because the group
velocity in a flat band is zero.}
because the eigensolutions are completely delocalized. Still the construction
of localized wave packets, consisting of eigensolutions peaked around a
specific momentum, is possible. These wave packets will spread due to the in
general nonlinear dispersion relation. Notice that now as opposed to the
classical wave equation there is no nonlinearity (the Schr\"odinger equation is
linear) that could cancel the effect of dispersion. Thus in accordance with the
uncertainty relation, the initial wave packet will spread. As is the case for
the free particle in quantum mechanics, the transition to classical mechanics
means that the spreading is going to zero.

Recent
work\cite{PhysRevA.80.053612,PhysRevLett.103.140403,Mishmash2009732,balakrishnan2009particle,satija2011other,reinhardt2011bright,rubbo2012quantum}
on the quantum dynamical aspects of solitons pursues the path of comparing
mean-field approximations with the quantum model. The mean-field
approach basically leads to a classical nonlinear equation of motion for
some operator expectation value restricted to a subset of carefully chosen
states (mainly product states). This classical nonlinear equation for an
operator expectation value, e.g., $\expect{\op{a}}$ for the condensate density
in a Bose-Einstein condensate, might exhibit soliton solutions. Its time
evolution is then compared for both the mean-field approximation as well for
the quantum evolution on the full Hilbert space. One intrinsic problem of this
approach is that it has to be justified that the mean-field approximation is
still valid for the time evolution and not just for its initial state.
We take a different route in that we identify directly a classical nonlinear
model with a quantum model using the direct canonical quantization. Therefore
we will get a one-to-one correspondence between classical soliton solutions
and their quantum mechanical counterparts or vice versa. Thus, in our
description, both the classical and the quantum model exhibit the same soliton
solutions and there is no intrinsic quantum soliton that does not exist in the
classical theory or the other way round. In that sense we define the term of a
quantum soliton as a state that would correspond up to the uncertainty
relation to a classical soliton state. It is an interesting question if there
exist models that exhibit intrinsic quantum solitons which do not occur in the
corresponding classical theory. But before answering this question, a scheme
for describing or defining an intrinsic quantum soliton must be found.

The simulation of real time dynamics in quantum systems is a numerically hard
problem due to the exponential increase of the Hilbert space with system size.
The development of the DMRG\cite{PhysRevLett.69.2863,Ulrich201196} algorithm
and its real time variant
t-DMRG\cite{PhysRevLett.91.147902,daley2004time,Ulrich201196} has opened up new
perspectives in simulating 1-d systems, whose size is far beyond of those that
are reachable by ED (see for example Refs.\
\onlinecite{PhysRevE.71.036102,PhysRevB.79.155104,PhysRevA.80.041603,rubbo2012quantum,PhysRevA.80.053612}). Both
methods allow us to create wave packets for larger systems.

In the following we will show that quantum wave packets can be constructed,
whose time-evolution is in agreement (despite the quantum mechanical spreading)
with their classical soliton counterparts. 

\section{The model}
To investigate the correspondence between classical and quantum mechanical
solitons, we use a 1-d spin chain as our model. For the ferromagnetic easy axis
spin chain, as described in the following paragraphs, exact solutions for the
energy spectra of the low-lying solutions exist. Both models are integrable,
the classical in the sense of the inverse scattering transform\cite{gardner1967method}
and the $s = \nicefrac{1}{2}$ quantum model in the sense of the
Bethe Ansatz.\cite{bethe1931theorie} Furthermore, a clear mapping between both
models exists, and thus facilitate the comparison between classical and quantum
eigensolutions and their spectra. %
\subsection{Quantum model}
The quantum model is described by the anisotropic Heisenberg Hamiltonian:
\begin{equation}
  \op{H} = -J \sum_i \left[ \frac{1}{2}\left( \op{S}^+_i \op{S}^-_{i+1} +
  \op{S}^+_{i+1} \op{S}^-_i \right) + \Delta \op{S}^z_i \op{S}^z_{i+1} \right]
  \, .
  \label{eqn:heis_aniso}
\end{equation}
We assume a ferromagnetic coupling $J > 0$, easy axis anisotropy $1 < \Delta =
1 + \Delta_z = \cosh \Phi$. Our reason for taking easy axis anisotropy ($1 <
\Delta$) is that in this case the analytical treatment of both the quantum and
the classical model is simplest.\cite{[][{ [\href{http://www.jetpletters.ac.ru/ps/1644/article_25093.shtml}{JETP Lett. \textbf{5}, 38
(1967)}].}]ovchinnikov1967complexes,kosevich1990magnetic} %
The ground state of this model is twofold degenerate (all spins pointing up or
down). In the following we will take the state $\ket{\!\!\downarrow\downarrow ..
\downarrow\downarrow}$ as the reference ground state.
Note that the $s = \nicefrac{1}{2}$ version of Eq.~%
(\ref{eqn:heis_aniso}) is equivalent to a hard-core Bose-Hubbard
model.\cite{matsubara1956lattice} Consequently, each flipped spin with respect
to the references state can be interpreted as occupation by one boson.

The energy for the lowest lying excitations with momentum $-\pi \le k \le \pi$
and magnetization $m$ (number of flipped spins) for $s = \nicefrac{1}{2}$ is
given by%
\cite{ovchinnikov1967complexes}%
\begin{equation}
  E_m(k) = J \cdot \left( \cosh m \Phi - \cos k \right)
  \frac{\sinh \Phi}{\sinh m \Phi}
  \label{eqn:heis_disp}
\end{equation}
in the thermodynamical limit. These excitations are also called $m$-magnon
bound states and are completely delocalized over the whole system. To get
localized excitations (i.e., where the magnetization is distributed over a region
of a few sites), which could correspond to localized classical solitons, it is
necessary to construct wave packets. These wave packets will consist of
$m$-magnon bound states with different momenta. If this momentum distribution
is peaked around $k$, we would expect a group velocity given by the derivative
of (\ref{eqn:heis_disp}):
\begin{equation}
  v_G\left( m, k \right) = J \sin k \frac{\sinh \Phi}{\sinh m\Phi} \, .
  \label{eqn:heis_vg}
\end{equation}
The maximum velocity of these wave packets is hence given by:
\begin{equation}
  v_{\text{max}}(m) = v_G\left( m, \pm \frac{\pi}{2} \right) = 
  \pm J \frac{\sinh \Phi}{\sinh m\Phi} \, .
  \label{eqn:heis_vmax}
\end{equation}

\subsection{Classical model}
The classical model is described by the Landau-Lifshitz equation (LLE).%
\cite{landau1935theory}
It can be derived from (\ref{eqn:heis_aniso}) by two approximations.
\begin{compactenum}
  \item  $\vecop{S} \rightarrow \vect{S} = s \left( \sin \theta \cos \varphi,
    \sin \theta \sin \varphi, \cos \theta \right)^{\text{t}}$, the classical
    treatment of spins which leads to an error of order $\nicefrac{1}{s}$.
  \item $\vect{S}_i \rightarrow \vect{S}(x_i)$, the continuum treatment via the
		long wavelength approximation with an error of order $\Delta_z^2$ for low
		lying excitations.
\end{compactenum}
The classical Hamiltonian is then given by
\begin{equation}
  H = J \int \text{d}x \left[ \frac{1}{2} \left( \frac{\partial
  \vect{S}}{\partial x} \right)^2 + \Delta_z \left( s^2 - S^2_z \right) \right]
  \, .
  \label{eqn:lle_ham}
\end{equation}
For the low-energy excitations of this Hamiltonian (1-soliton solutions), the
quasi-classical quantization%
\footnote{That is, identifying classical magnetization and momentum with their
quantum mechanical counter-part.}
gives an energy dispersion\cite{kosevich1977nonlinear}
\begin{equation}
  E(m, k) = 4s^2 J \sqrt{2\Delta_z}\left( 
  \frac{\cosh m\sqrt{2\Delta_z}/2s - \cos k}{\sinh m\sqrt{2\Delta_z}/2s}\right).
  \label{eqn:lle_disp}
\end{equation}
It can be seen that for $s = \nicefrac{1}{2}$ (\ref{eqn:lle_disp}) is exactly
the same as (\ref{eqn:heis_disp}) to first order in $\Delta_z$. Hence, even for
the ``most'' quantum-like case ($s = \nicefrac{1}{2}$), the energy spectra of
the low-energy excitations are identical for both the quantum and the classical
model.
This leads naturally to an identification between classical and quantum
solutions. But it is anyway an oddity that the classical spin profile
$\vect{S}(x)$ is localized in space and the quantum profile $\expect{\op{S}_i}$
is completely delocalized. So in order to get a classical profile coming from
the quantum model, it seems natural to build wave packets as in the well known
problem of a free particle.

\section{Numerical Investigations}
In the following simulations we assume $J = 1$ and $s = \nicefrac{1}{2}$. The
DMRG and t-DMRG algorithms were used for Figs.~\ref{fig:flip_three} and
\ref{fig:inter} using open boundary conditions (OBC) and a discarded weight of
$10^{-9}$ for the time evolutions.%
\footnote{It should be noted that the time evolution in
Figs.~\ref{fig:flip_three}, \ref{fig:inter}(b) and \ref{fig:inter}(c) could
have also been done using ED instead of t-DMRG.}
Exact diagonalization for calculating the time evolution was used in Figs.~%
\ref{fig:imp0}, \ref{fig:1imprint} and \ref{fig:2imprint}. Here, periodic
boundary conditions (PBC) were used in order to calculate the weights in Eq.~%
(\ref{eqn:mom_weight}) below in the basis of momentum eigenstates. For both
methods a second order Trotter decomposition\cite{suzuki1976generalized} with
$\Delta t = 0.01$ was used for time evolution.

\subsection{Single spin flips}
\begin{figure}[t!]
  \includegraphics[width=\linewidth]{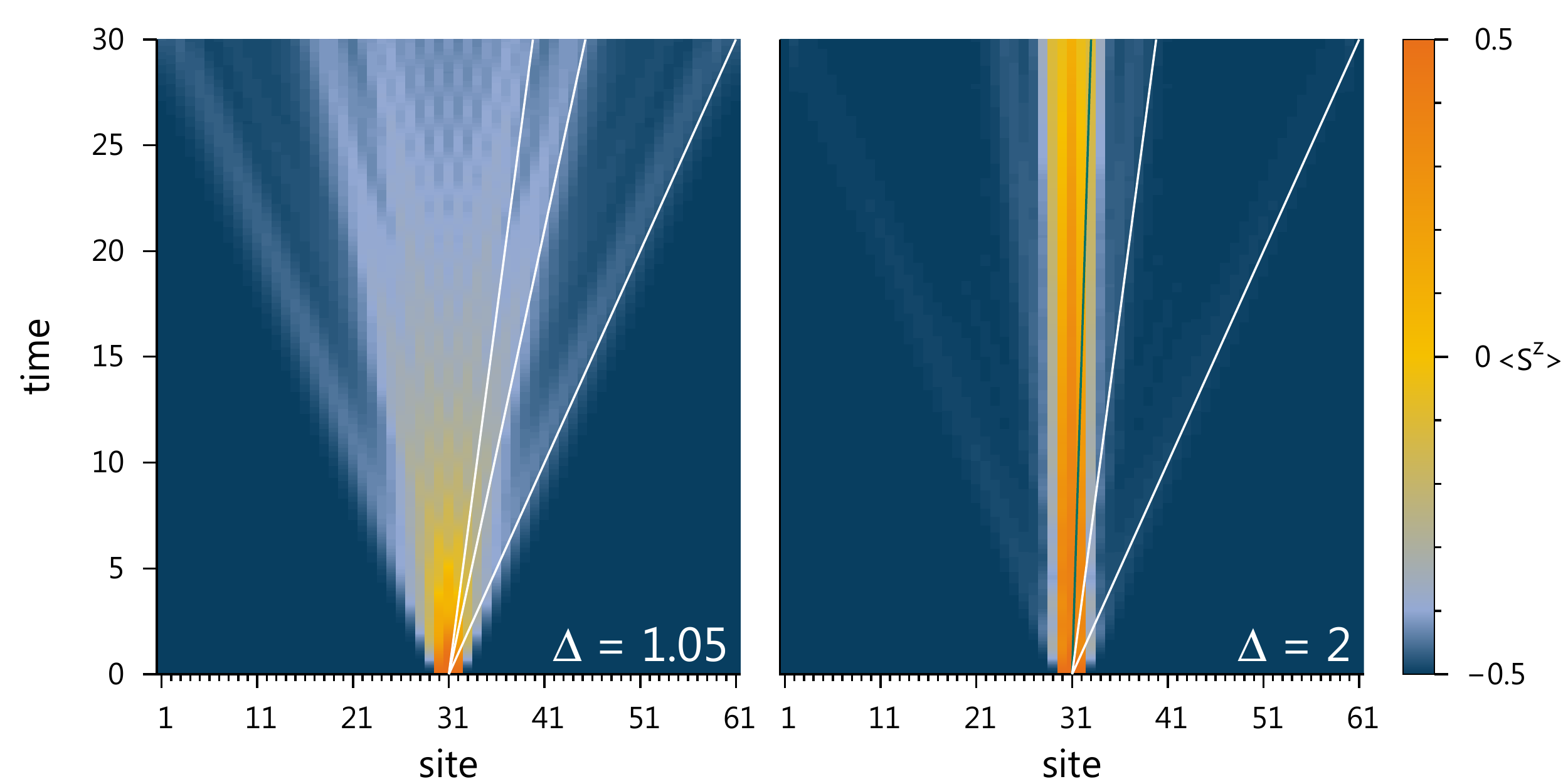}%
  \caption{(Color online)
  \label{fig:flip_three} Time evolution of Fock state
  $\ket{\!\!\downarrow \downarrow .. \downarrow \uparrow \uparrow \uparrow
  \downarrow .. \downarrow \downarrow}$ for $s = \nicefrac{1}{2}$,
  $\Delta = 1.05$ (left) and $\Delta
  = 2$ (right). Lines show the expected movement with maximum group velocity
  $v_{\text{max}}$ (see (\ref{eqn:heis_vmax})).}
\end{figure}
A very crude way to create wave packets is by flipping single spins from the
ground state. This will lead to a very broad distribution in momentum space,
meaning that excitations with different momenta and magnetization will be
created. This can be seen clearly in Fig.~\ref{fig:flip_three}.  Three spins
are flipped in the middle of the chain. The lines correspond to excitations
with maximum group velocity $v_{\text{max}}$ according to (\ref{eqn:heis_vmax})
for different $m$ ($= 3, 2, 1$ from left to right) and $\Delta$. Thus, it can
be seen that also 1-magnon (spin waves) and 2-magnon bound states are excited
by a simple 3-spin flip. This is consistent, because for $m = 3$ there exist
higher excitations consisting of $m = 2$ bound states plus one $m = 1$
scattering state as well as 3 times $m = 1$ scattering states. From the point
of view of classic integrability, meaning excitations will go through each
other without interaction, this dissection of the spectra is also necessary.
If, for example, our initial state consists of 2 $\times$ 3-magnon wave
packets, these will not disperse into 3 $\times$ 2-magnon wave packets during a
collision because the time evolution will always stay in the initial 2 $\times$
3-magnon bound-state sector.

\subsection{Constructing wave packets}
The following way to create specific excitations is based on the ideas of
Ref.~\onlinecite{PhysRevA.63.051601} to create dark solitons in Bose-Einstein
condensates. Our scheme is very similar and consists of three main steps:

1. Instead of flipping $m$ spins in the middle of the chain, a more delocalized
(in real space) wave packet will be created by adding a magnetic field
$B_{\text{loc}}$ to the Hamiltonian (\ref{eqn:heis_aniso}) which shall attract
the flipped spins to the middle. Whence, numerical methods will yield a
ground state where the flipped spins (whose number can be set by the initial
magnetization) rest at the center. 

2. Because of the symmetry, in momentum space the wave packet will be localized
around $k = 0$. In order to kick this wave packet an additional time evolution
is done just with a specific magnetic field $B_{\text{phase}}$.

3. After this initial preparation of the wave packet, its free time evolution
under the Hamiltonian (\ref{eqn:heis_aniso}) can be investigated.

Previous numerical work using this method has been done in
Refs.~\onlinecite{PhysRevA.80.053612,PhysRevLett.103.140403,Mishmash2009732}.
\subsubsection{Localizing the wave packet}
\begin{figure}[t!]
  \includegraphics[width=\linewidth]{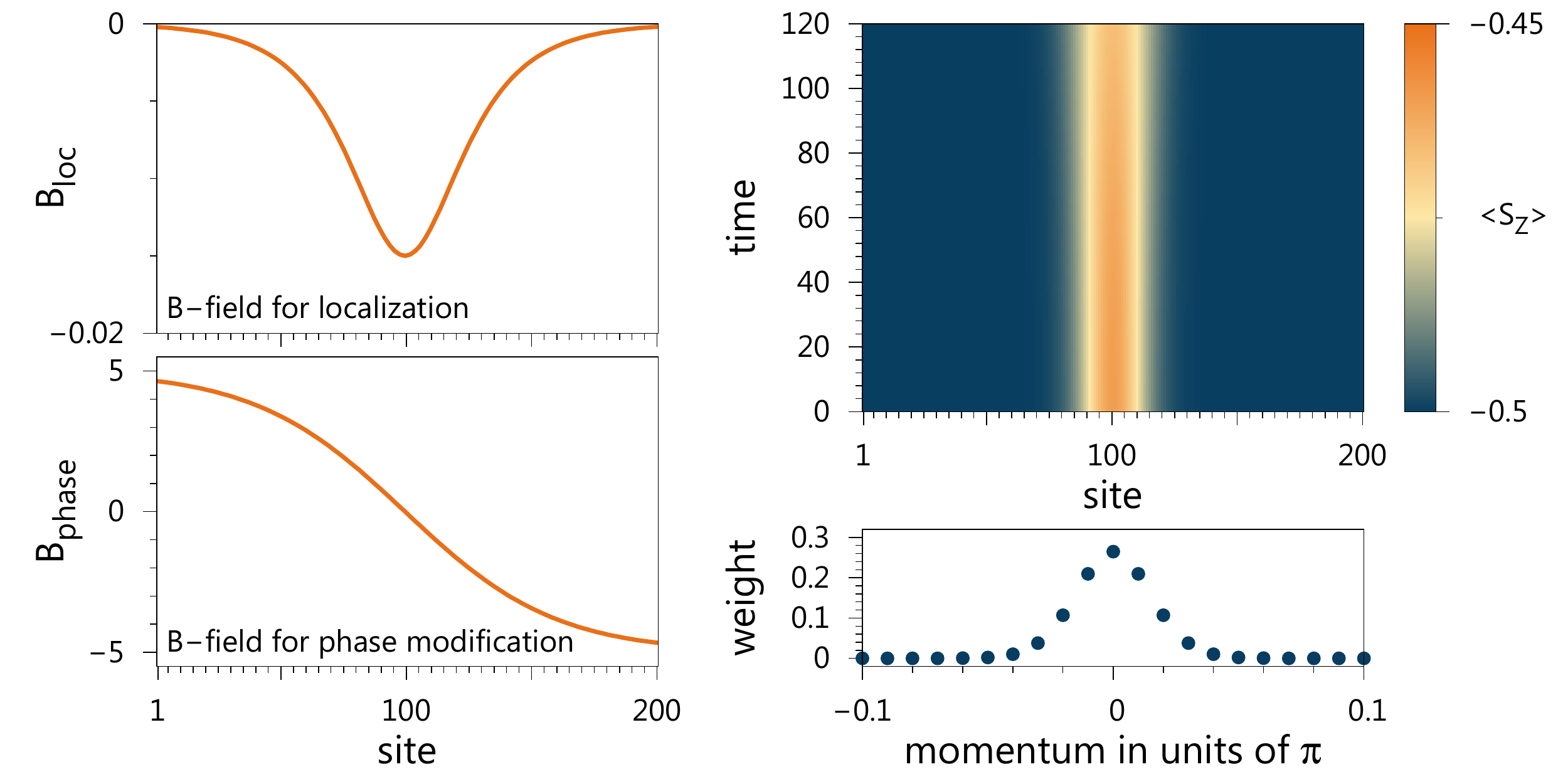}%
  \caption{(Color online)
  \label{fig:imp0} (\textbf{left}): Schematic profile of the magnetic
  fields for localizing the excitation and for shifting its momentum
  distribution.  (\textbf{right}): Time evolution of a localized excitation ($m
  = 1$, $\Delta = 1.05$) and its momentum distribution before applying the
  $\op{H}_{\text{phase}}$. Its initial width is bigger than in
  Fig.~\ref{fig:flip_three} but spreads only slightly because of the sharp
  momentum distribution around $k = 0$.}
\end{figure}
Because a simple spin flip creates a completely localized excitation, the
momentum distribution will be completely smeared out and the localized
excitation will dislocate very quickly. To get an initial state, which is also
localized in momentum space, it is thus necessary to have some delocalization
in real space. To create such a state, we add a magnetic field for localization
(see Fig.~\ref{fig:imp0}) to the Hamiltonian (\ref{eqn:heis_aniso}) of the
following form:
\begin{eqnarray}
  \label{eqn:hloc}
  \op{H}_{\text{loc}} & = & \sum_i \frac{1}{s} B_{\text{loc}}[i] \cdot \op{S}^z_i \, , \\
  \text{with} \quad B_{\text{loc}}[i] & = & -\frac{B_{\text{LocA}}}{\cosh \left( \frac{x_0 - i}{B_{\text{LocW}}} \right)} \, . \nonumber
\end{eqnarray}
Fixing the magnetization $m$ (i.e., the number of flipped spins) and calculating
the ground state will result in a magnetization profile as can be seen in the
right part of Fig.~\ref{fig:imp0}. The term $\op{H}_{\text{loc}}$ is only used
for the initial state. Time evolution is done just with (\ref{eqn:heis_aniso}).
The parameters $B_{\text{LocA}}$ and $B_{\text{LocW}}$ control the depth and
width of the magnetic field and therefore the localization of the wave packet.

Using exact diagonalization in momentum space, the projection of the initial
state to the momentum eigenstates of (\ref{eqn:heis_aniso}) can be calculated
as well as their weight
\begin{equation}
  \text{weight}\left( k, \alpha \right) =
  \norm{\scalprod{\psi_{\text{excited}}}{k_{\alpha}}} \, .
  \label{eqn:mom_weight}
\end{equation}
The index $\alpha$ runs through the number of eigenstates with momentum $k$.
For $m = 1$, there is just one such eigenstate for each $k$. This weight
distribution is shown below the time evolutions in Fig.~\ref{fig:imp0} and
\ref{fig:1imprint}. The peaked momentum distribution around $k = 0$ in
Fig.~\ref{fig:imp0} clarifies the stability  of the magnetization profile.
\subsubsection{Kicking the wave packet}
\begin{figure}[h!]
  \includegraphics[width=\linewidth]{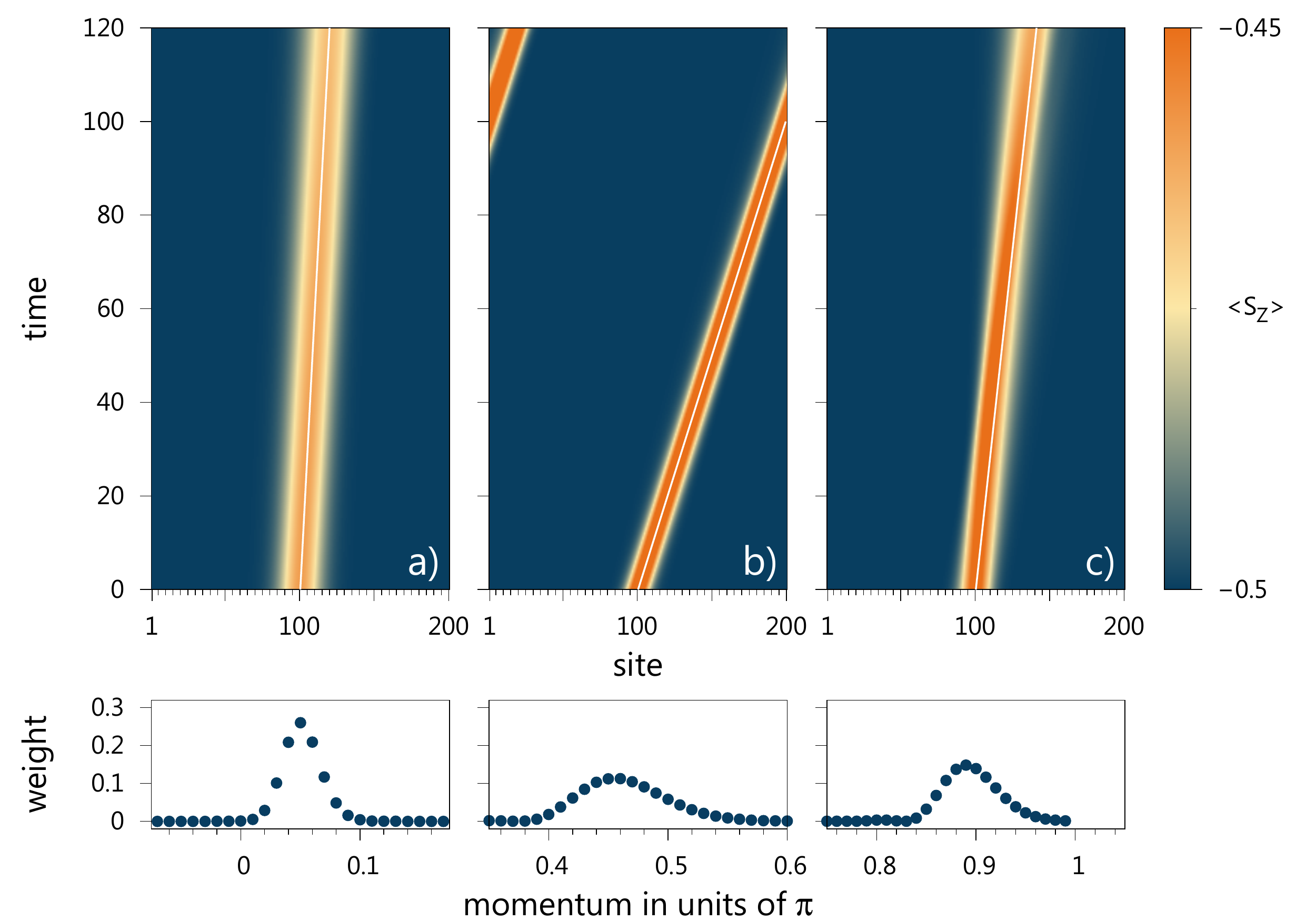}%
  \caption{(Color online)
  \label{fig:1imprint} Time evolution for $m = 1$, $\Delta = 1.05$
  using different $\op{H}_{\text{loc}}$ and $\op{H}_{\text{phase}}$ (see Table
  \ref{tab:imprint}) and its momentum distribution. White lines show the
  expected movement with group velocity $v_g(k_P)$, where $k_P$ is the momentum
  with maximum weight.}
\end{figure}
To get the localized wave packet into movement it is necessary to shift the
momentum distribution to a $k \ne 0$. Changing the phase of each Fock state in
real space that the initial state consists of will not change the magnetization
profile but the momentum distribution. This phase change can be
implemented by a time evolution of the initial state with the following
Hamiltonian:
\begin{eqnarray}
  \label{eqn:hphase}
  \op{H}_{\text{phase}} & = & \sum_i \frac{1}{s} B_{\text{phase}}[i] \cdot 
  \op{S}^z_i \, , \\
  \text{with} \quad B_{\text{phase}}[i] & = & B_{\text{PhA}} \tanh \left( 
  \frac{x_0 - i}{B_{\text{PhW}}}\right) \, .\nonumber
\end{eqnarray}
The magnetic field $B_{\text{phase}}[i]$ (sketched in Fig.~\ref{fig:imp0})
would also suggest, that the wave packet would slide down to the right
corresponding to a momentum shift to $k > 0$. This is indeed the case as can be
seen in Fig.~\ref{fig:1imprint}. $B_{\text{PhA}}$ and $B_{\text{PhW}}$ modify
the amplitude and width of the phase imprinting magnetic field. Concerning the
length $t_{\text{phase}}$ of the time evolution with $\op{H}_{\text{phase}}$ it
should be noted that the phase imprinted state depends only on the product
$t_{\text{phase}} \cdot B_{\text{PhA}}$. That is why we fixed $t_{\text{phase}}
= 1$ and varied $B_{\text{PhA}}$. Fig.~\ref{fig:1imprint} shows thus that
the picture of moving wave packets with specific group velocity (defined by the
peak in their momentum distribution) is consistent. The parameters for
$\op{H}_{\text{loc}}$ and $\op{H}_{\text{phase}}$ were found%
\footnote{It should be mentioned that these wave packets can be created
directly by a superposition of the specific momentum eigenstate, if these are
known analytically.}
by trial \& error
and are given in Table \ref{tab:imprint}.
\begin{figure}[t!]
  \includegraphics[width=\linewidth]{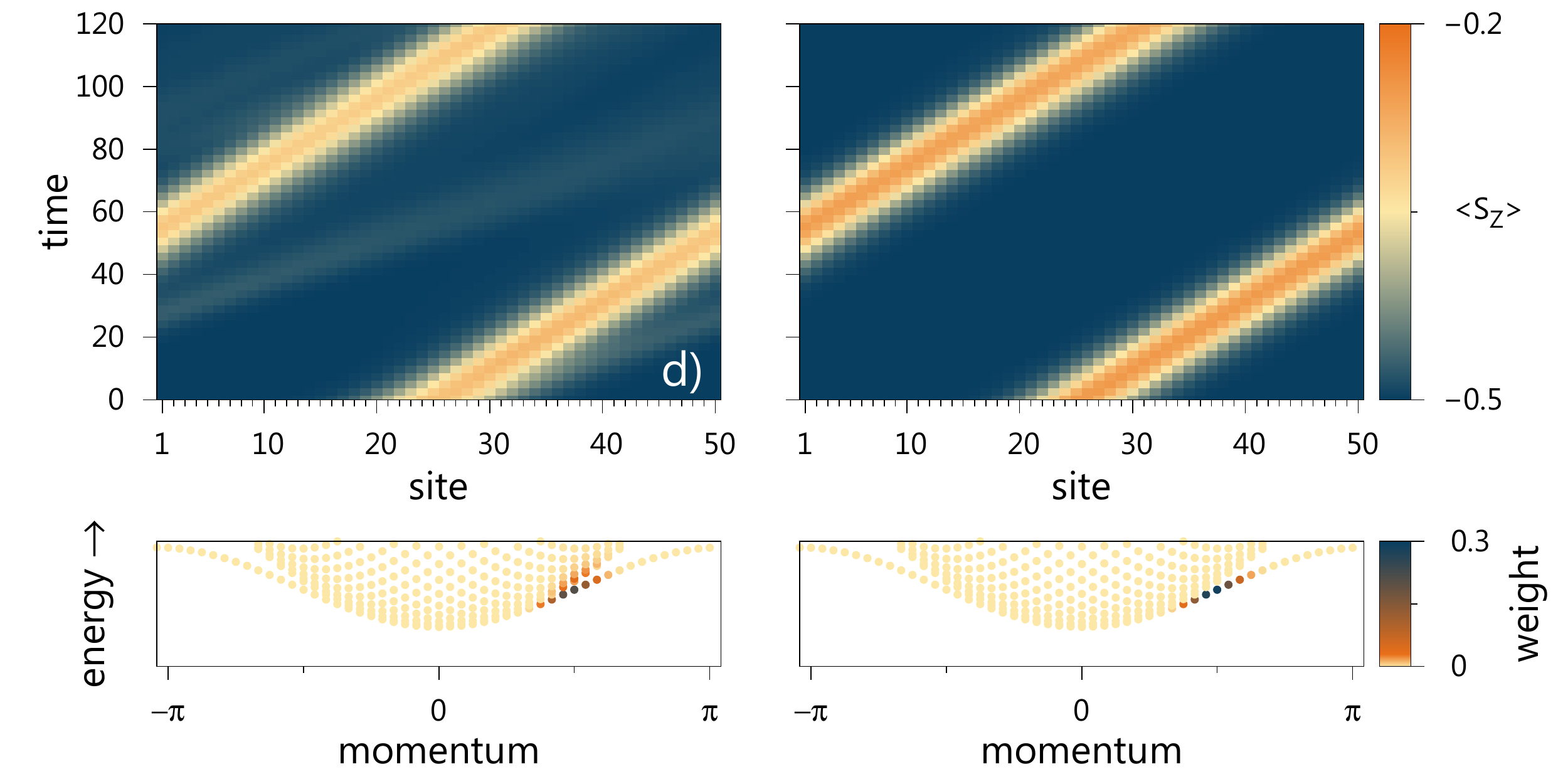}%
  \caption{(Color online)
  \label{fig:2imprint} (\textbf{left}): Time evolution for $m = 2$,
  $\Delta = 1.05$ using $\op{H}_{\text{loc}}$ and $\op{H}_{\text{phase}}$ as
  specified for case d) in Table \ref{tab:imprint}. A faster ray right to the
  main $m = 2$ moving bound excitation can be seen, which is attributed to the
  weight in the $m = 1 + 1$ scattering sector of the momentum distribution.
  (\textbf{right}): By projecting out the states with the weight in the $m = 1
  + 1$ scattering sector, the faster ray disappears.}
\end{figure}
\squeezetable
\begin{table}[t!]
  \begin{ruledtabular}
    \begin{tabular}{@{}rh{4}h{0}h{1}h{0}@{}}
      Case &
      \multicolumn{1}{c}{$B_{\text{LocA}}$} & 
      \multicolumn{1}{c}{$B_{\text{LocW}}$} &
      \multicolumn{1}{c}{$B_{\text{PhA}}$} &
      \multicolumn{1}{c}{$B_{\text{PhW}}$} \\
      \hline
      \\[-6pt]
      a) & 0.0075 & 20 & 5 & 60 \\
      b) & 0.02 & 7 & 22.5 & 30 \\
      c) & 0.15 & 50 & 85.5 & 60 \\
      d) & 0.05 & 10 & 15 & 40 \\
    \end{tabular}
  \end{ruledtabular}
  \caption{\label{tab:imprint} Parameters for $\op{H}_{\text{loc}}$ and
  $\op{H}_{\text{phase}}$ to generate the excitations shown in
  Figs.~\ref{fig:1imprint}--\ref{fig:2imprint}.}
\end{table}

Using this scheme for creating wave packets with $m > 1$ will of course result
also in other higher excitations. This is shown in Fig.~\ref{fig:2imprint}
for $m = 2$. The slower moving excitation in the left part corresponds to a
2-bound wave packet, while the faster light one corresponds to a spin wave. The
distribution in momentum space also shows this. By using exact diagonalization
in momentum space it is possible to project out these spin wave excitations
from the 2 times $m = 1$ sector of the excitation spectra. The result of this
projection (the disappearance of the faster spin wave excitation) is seen in
the right part of Fig.~\ref{fig:2imprint}.

\subsubsection{Colliding wave packets}
\begin{figure}[t!]
  \includegraphics[width=\linewidth]{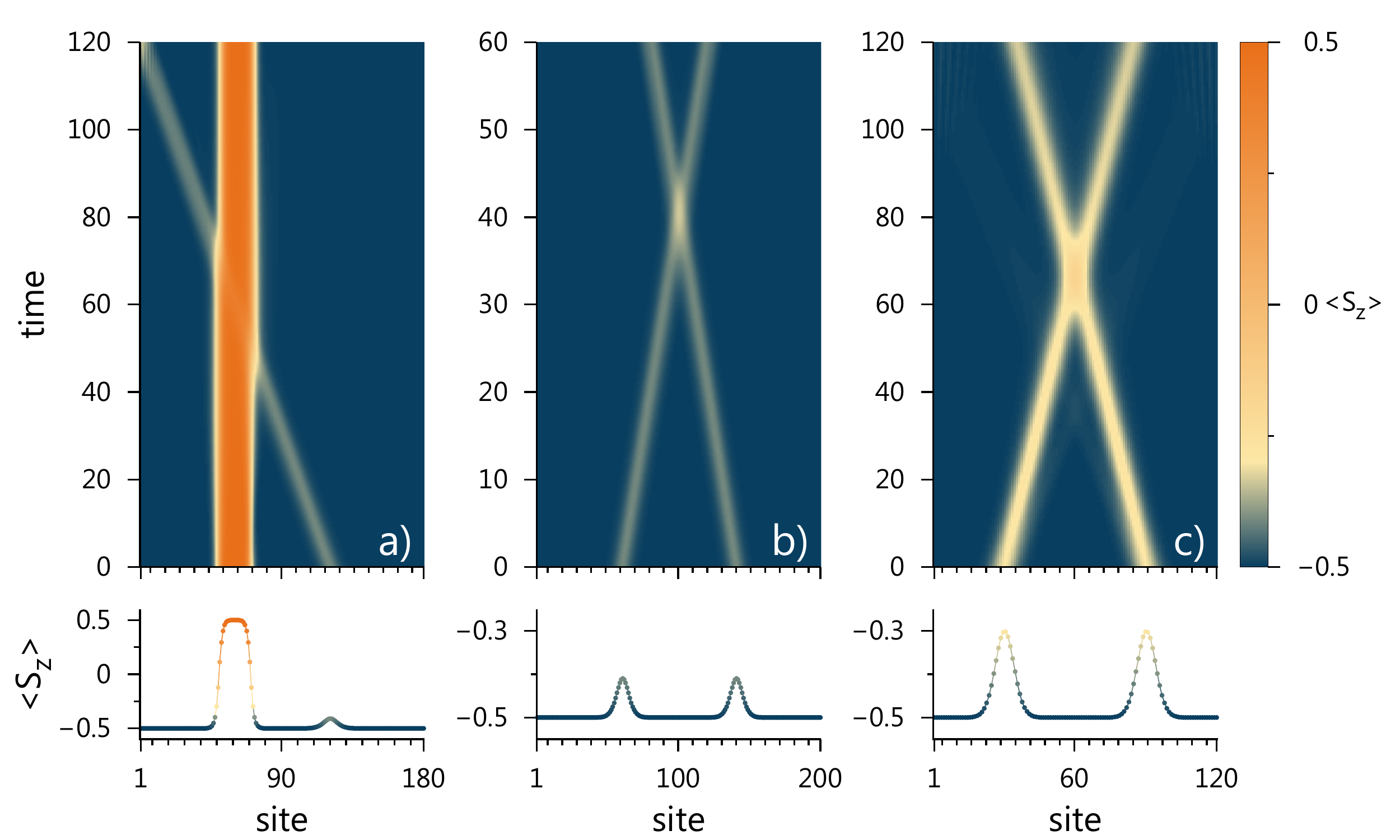}%
  \caption{(Color online)
  \label{fig:inter}(\textbf{top}): Time evolution of different
  combined excitations ($\Delta = 1.05$, for other parameters see Table
  \ref{tab:inter}). \textbf{a}) spin wave packet ($m = 1$) running through a
  large static magnon complex consisting of two bound domain walls ($m = 20$)
  and shifting it by 1 site to the right.
  \textbf{b}) 2 spin wave packets ($m = 1$) running through each other.
  \textbf{c}) 2 magnon packets ($m = 2$) running through each other.
  (\textbf{bottom}): corresponding initial magnetization profile}
\end{figure}
Using the methods described before it is also possible to create two wave
packets on a chain.%
\footnote{In order to create
more than one localized wave packet, it is necessary to remove
the interactions (spin hopping) between the parts of the chain, where these
excitations should be placed. Otherwise all the magnetization will fall into
one of the valleys of the magnetic field potential in the ground state.}
Various scenarios can be obtained this way. Fig.~\ref{fig:inter} shows 3
types of collisions. The first (a) shows the passing of a $m=1$ spin wave
through a resting $m=20$ bound state (which can be considered as two bound
domain walls). The movement of the domain wall by exactly one site
(corresponding to one unit of magnetization $m = 1$) was predicted earlier in
Ref.~\onlinecite{[{}][{ [\href{http://www.jetpletters.ac.ru/ps/1299/article_19617.shtml}{JETP Lett. \textbf{39}, 354 (1984)}].}]mikhailov1984forced} and can
be seen here. The other two settings (b) and (c) show the collision of two $m =
1$ and $m = 2$ wave packets, propagating in opposite directions. As known from
classical integrability, these excitations should just go through each other,
because their characteristics (i.e.,  momentum and magnetization) represent
integrals of motion. The parameters used for these calculations are given in
Table \ref{tab:inter}. 
\begin{table}[t!]
  \begin{ruledtabular}
    \begin{tabular}{@{}rlh{3}h{0}h{0}h{0}@{}}
      Case & &
      \multicolumn{1}{c}{$B_{\text{LocA}}$} & 
      \multicolumn{1}{c}{$B_{\text{LocW}}$} &
      \multicolumn{1}{c}{$B_{\text{PhA}}$} &
      \multicolumn{1}{c}{$B_{\text{PhW}}$} \\
      \hline \\[-6pt]
      a)& left: $N_M = 20$, $x_0 = 60$ & 0.05 & 5 & 0 & 25 \\
	& right: $N_M = 1$, $x_0 = 120$ & 0.02 & 5 & -20 & 25 \\
      b)& left: $N_M = 1$, $x_0 = 30$ & 0.02 & 5 & 20 & 25 \\
	& right: $N_M = 1$, $x_0 = 70$ & 0.02 & 5 & -20 & 25 \\
      c)& left: $N_M = 2$, $x_0 = 30$ & 0.05 & 10 & 15 & 40 \\
	& right: $N_M = 2$, $x_0 = 90$ & 0.05 & 10 & -15 & 40 \\
    \end{tabular}
  \end{ruledtabular}
  \caption{\label{tab:inter} Parameters for $\op{H}_{\text{loc}}$ and
  $\op{H}_{\text{phase}}$ to generate the excitations shown in
  Fig.~\ref{fig:inter}.}
\end{table}

\section{Conclusion} 
Based on the analytical findings of Ref.~\onlinecite{kosevich1990magnetic}, we
investigated the real-time evolution of quantum wave packets in the
ferromagnetic easy-axis Heisenberg model. They were constructed in a way close
to their classical soliton counterparts by using additional magnetic fields to
localize them and to give them a momentum kick. The time evolution is
consistent with the classical picture of the integrable LLE and in the case of
the setting shown in Fig.~\ref{fig:inter}(a) explicitly shows the analytical
predictions of Ref.~\onlinecite{mikhailov1984forced}.

This method of constructing localized wave packets might also be used for
non-integrable quantum systems, where colliding wave packets might excite each
other or slow down and create new excitations from the background. Furthermore
it might be used for examining transport properties in spin system.

\begin{acknowledgments}
  We thank H.~J.~Mikeska and A.~K.~Kolezhuk for useful discussions. A.W.~would like to
  thank R.~Peters and P.~E.~Dargel for support on the DMRG code. Parts of the
  DMRG code were based on prior work by R.~Peters. A.H.\ acknowledges
  support by the
  Deutsche Forschungsgemeinschaft via a Heisenberg fellowship (Project
  HO~2325/4-2).
\end{acknowledgments}

\bibliography{refs}

\begin{thebibliography}{36}%
\makeatletter
\providecommand \@ifxundefined [1]{%
 \@ifx{#1\undefined}
}%
\providecommand \@ifnum [1]{%
 \ifnum #1\expandafter \@firstoftwo
 \else \expandafter \@secondoftwo
 \fi
}%
\providecommand \@ifx [1]{%
 \ifx #1\expandafter \@firstoftwo
 \else \expandafter \@secondoftwo
 \fi
}%
\providecommand \natexlab [1]{#1}%
\providecommand \enquote  [1]{``#1''}%
\providecommand \bibnamefont  [1]{#1}%
\providecommand \bibfnamefont [1]{#1}%
\providecommand \citenamefont [1]{#1}%
\providecommand \href@noop [0]{\@secondoftwo}%
\providecommand \href [0]{\begingroup \@sanitize@url \@href}%
\providecommand \@href[1]{\@@startlink{#1}\@@href}%
\providecommand \@@href[1]{\endgroup#1\@@endlink}%
\providecommand \@sanitize@url [0]{\catcode `\\12\catcode `\$12\catcode
  `\&12\catcode `\#12\catcode `\^12\catcode `\_12\catcode `\%12\relax}%
\providecommand \@@startlink[1]{}%
\providecommand \@@endlink[0]{}%
\providecommand \url  [0]{\begingroup\@sanitize@url \@url }%
\providecommand \@url [1]{\endgroup\@href {#1}{\urlprefix }}%
\providecommand \urlprefix  [0]{URL }%
\providecommand \Eprint [0]{\href }%
\providecommand \doibase [0]{http://dx.doi.org/}%
\providecommand \selectlanguage [0]{\@gobble}%
\providecommand \bibinfo  [0]{\@secondoftwo}%
\providecommand \bibfield  [0]{\@secondoftwo}%
\providecommand \translation [1]{[#1]}%
\providecommand \BibitemOpen [0]{}%
\providecommand \bibitemStop [0]{}%
\providecommand \bibitemNoStop [0]{.\EOS\space}%
\providecommand \EOS [0]{\spacefactor3000\relax}%
\providecommand \BibitemShut  [1]{\csname bibitem#1\endcsname}%
\let\auto@bib@innerbib\@empty
\bibitem [{\citenamefont {Scott}\ \emph {et~al.}(1973)\citenamefont {Scott},
  \citenamefont {Chu},\ and\ \citenamefont {McLaughlin}}]{scott1973soliton}%
  \BibitemOpen
  \bibfield  {author} {\bibinfo {author} {\bibfnamefont {A.~C.}\ \bibnamefont
  {Scott}}, \bibinfo {author} {\bibfnamefont {F.~Y.~F.}\ \bibnamefont {Chu}}, \
  and\ \bibinfo {author} {\bibfnamefont {D.~W.}\ \bibnamefont {McLaughlin}},\
  }\href@noop {} {\bibfield  {journal} {\bibinfo  {journal} {Proc. IEEE}\
  }\textbf {\bibinfo {volume} {61}},\ \bibinfo {pages} {1443} (\bibinfo {year}
  {1973})}\BibitemShut {NoStop}%
\bibitem [{\citenamefont {Zabusky}\ and\ \citenamefont
  {Kruskal}(1965)}]{zabusky1965interaction}%
  \BibitemOpen
  \bibfield  {author} {\bibinfo {author} {\bibfnamefont {N.~J.}\ \bibnamefont
  {Zabusky}}\ and\ \bibinfo {author} {\bibfnamefont {M.~D.}\ \bibnamefont
  {Kruskal}},\ }\href@noop {} {\bibfield  {journal} {\bibinfo  {journal} {Phys.
  Rev. Lett.}\ }\textbf {\bibinfo {volume} {15}},\ \bibinfo {pages} {240}
  (\bibinfo {year} {1965})}\BibitemShut {NoStop}%
\bibitem [{\citenamefont {Gardner}\ \emph {et~al.}(1967)\citenamefont
  {Gardner}, \citenamefont {Greene}, \citenamefont {Kruskal},\ and\
  \citenamefont {Miura}}]{gardner1967method}%
  \BibitemOpen
  \bibfield  {author} {\bibinfo {author} {\bibfnamefont {C.~S.}\ \bibnamefont
  {Gardner}}, \bibinfo {author} {\bibfnamefont {J.~M.}\ \bibnamefont {Greene}},
  \bibinfo {author} {\bibfnamefont {M.~D.}\ \bibnamefont {Kruskal}}, \ and\
  \bibinfo {author} {\bibfnamefont {R.~M.}\ \bibnamefont {Miura}},\ }\href@noop
  {} {\bibfield  {journal} {\bibinfo  {journal} {Phys. Rev. Lett.}\ }\textbf
  {\bibinfo {volume} {19}},\ \bibinfo {pages} {1095} (\bibinfo {year}
  {1967})}\BibitemShut {NoStop}%
\bibitem [{Note1()}]{Note1}%
  \BibitemOpen
  \bibinfo {note} {The term ``integrable'' is well defined for classical
  mechanics. As for quantum mechanics, we refer to this term in the sense of
  ``soluble by the quantum inverse scattering method'' (see Ref.~\protect
  \rev@citealpnum {korepin1997quantum}).}\BibitemShut {Stop}%
\bibitem [{\citenamefont {Zakharov}\ and\ \citenamefont
  {Faddeev}(1971)}]{zakharov1971korteweg}%
  \BibitemOpen
  \bibfield  {author} {\bibinfo {author} {\bibfnamefont {V.~E.}\ \bibnamefont
  {Zakharov}}\ and\ \bibinfo {author} {\bibfnamefont {L.~D.}\ \bibnamefont
  {Faddeev}},\ }\href@noop {} {\bibfield  {journal} {\bibinfo  {journal}
  {Funct. Anal. Appl.}\ }\textbf {\bibinfo {volume} {5}},\ \bibinfo {pages}
  {280} (\bibinfo {year} {1971})}\BibitemShut {NoStop}%
\bibitem [{\citenamefont {Kosevich}\ \emph {et~al.}(1977)\citenamefont
  {Kosevich}, \citenamefont {Ivanov},\ and\ \citenamefont
  {Kovalev}}]{kosevich1977nonlinear}%
  \BibitemOpen
  \bibfield  {author} {\bibinfo {author} {\bibfnamefont {A.~M.}\ \bibnamefont
  {Kosevich}}, \bibinfo {author} {\bibfnamefont {B.~A.}\ \bibnamefont
  {Ivanov}}, \ and\ \bibinfo {author} {\bibfnamefont {A.~S.}\ \bibnamefont
  {Kovalev}},\ }\href@noop {} {\bibfield  {journal} {\bibinfo  {journal} {Fiz.
  Nizk. Temp.}\ }\textbf {\bibinfo {volume} {3}},\ \bibinfo {pages} {906}
  (\bibinfo {year} {1977})}\BibitemShut {NoStop}%
\bibitem [{Note2()}]{Note2}%
  \BibitemOpen
  \bibinfo {note} {Localized eigenstates for translationally invariant quantum
  models might exist if the model has a flat band (as for quite a few
  frustrated models\cite {[{}][{ [Fizika Nizkikh Temperatur \protect \textbf
  {33}, 982 (2007)].}]derzhko:745,PhysRevB.82.184502}), but for the quantized
  versions of the nonlinear wave equations mentioned in this work, this is not
  the case. Furthermore these types of localized eigenstates cannot move
  because the group velocity in a flat band is zero.}\BibitemShut {Stop}%
\bibitem [{\citenamefont {Mishmash}\ \emph {et~al.}(2009)\citenamefont
  {Mishmash}, \citenamefont {Danshita}, \citenamefont {Clark},\ and\
  \citenamefont {Carr}}]{PhysRevA.80.053612}%
  \BibitemOpen
  \bibfield  {author} {\bibinfo {author} {\bibfnamefont {R.~V.}\ \bibnamefont
  {Mishmash}}, \bibinfo {author} {\bibfnamefont {I.}~\bibnamefont {Danshita}},
  \bibinfo {author} {\bibfnamefont {C.~W.}\ \bibnamefont {Clark}}, \ and\
  \bibinfo {author} {\bibfnamefont {L.~D.}\ \bibnamefont {Carr}},\ }\href
  {\doibase 10.1103/PhysRevA.80.053612} {\bibfield  {journal} {\bibinfo
  {journal} {Phys. Rev. A}\ }\textbf {\bibinfo {volume} {80}},\ \bibinfo
  {pages} {053612} (\bibinfo {year} {2009})}\BibitemShut {NoStop}%
\bibitem [{\citenamefont {Mishmash}\ and\ \citenamefont
  {Carr}(2009{\natexlab{a}})}]{PhysRevLett.103.140403}%
  \BibitemOpen
  \bibfield  {author} {\bibinfo {author} {\bibfnamefont {R.~V.}\ \bibnamefont
  {Mishmash}}\ and\ \bibinfo {author} {\bibfnamefont {L.~D.}\ \bibnamefont
  {Carr}},\ }\href {\doibase 10.1103/PhysRevLett.103.140403} {\bibfield
  {journal} {\bibinfo  {journal} {Phys. Rev. Lett.}\ }\textbf {\bibinfo
  {volume} {103}},\ \bibinfo {pages} {140403} (\bibinfo {year}
  {2009}{\natexlab{a}})}\BibitemShut {NoStop}%
\bibitem [{\citenamefont {Mishmash}\ and\ \citenamefont
  {Carr}(2009{\natexlab{b}})}]{Mishmash2009732}%
  \BibitemOpen
  \bibfield  {author} {\bibinfo {author} {\bibfnamefont {R.~V.}\ \bibnamefont
  {Mishmash}}\ and\ \bibinfo {author} {\bibfnamefont {L.~D.}\ \bibnamefont
  {Carr}},\ }\href {\doibase 10.1016/j.matcom.2009.08.025} {\bibfield
  {journal} {\bibinfo  {journal} {Math. Comput. Simulat.}\ }\textbf {\bibinfo
  {volume} {80}},\ \bibinfo {pages} {732 } (\bibinfo {year}
  {2009}{\natexlab{b}})}\BibitemShut {NoStop}%
\bibitem [{\citenamefont {Balakrishnan}\ \emph {et~al.}(2009)\citenamefont
  {Balakrishnan}, \citenamefont {Satija},\ and\ \citenamefont
  {Clark}}]{balakrishnan2009particle}%
  \BibitemOpen
  \bibfield  {author} {\bibinfo {author} {\bibfnamefont {R.}~\bibnamefont
  {Balakrishnan}}, \bibinfo {author} {\bibfnamefont {I.~I.}\ \bibnamefont
  {Satija}}, \ and\ \bibinfo {author} {\bibfnamefont {C.~W.}\ \bibnamefont
  {Clark}},\ }\href@noop {} {\bibfield  {journal} {\bibinfo  {journal} {Phys.
  Rev. Lett.}\ }\textbf {\bibinfo {volume} {103}},\ \bibinfo {pages} {230403}
  (\bibinfo {year} {2009})}\BibitemShut {NoStop}%
\bibitem [{\citenamefont {Satija}\ and\ \citenamefont
  {Balakrishnan}(2011)}]{satija2011other}%
  \BibitemOpen
  \bibfield  {author} {\bibinfo {author} {\bibfnamefont {I.~I.}\ \bibnamefont
  {Satija}}\ and\ \bibinfo {author} {\bibfnamefont {R.}~\bibnamefont
  {Balakrishnan}},\ }\href@noop {} {\bibfield  {journal} {\bibinfo  {journal}
  {Phys. Lett. A}\ }\textbf {\bibinfo {volume} {375}},\ \bibinfo {pages} {517}
  (\bibinfo {year} {2011})}\BibitemShut {NoStop}%
\bibitem [{\citenamefont {Reinhardt}\ \emph {et~al.}(2011)\citenamefont
  {Reinhardt}, \citenamefont {Satija}, \citenamefont {Robbins},\ and\
  \citenamefont {Clark}}]{reinhardt2011bright}%
  \BibitemOpen
  \bibfield  {author} {\bibinfo {author} {\bibfnamefont {W.~P.}\ \bibnamefont
  {Reinhardt}}, \bibinfo {author} {\bibfnamefont {I.~I.}\ \bibnamefont
  {Satija}}, \bibinfo {author} {\bibfnamefont {B.}~\bibnamefont {Robbins}}, \
  and\ \bibinfo {author} {\bibfnamefont {C.~W.}\ \bibnamefont {Clark}},\
  }\href@noop {} {\bibfield  {journal} {\bibinfo  {journal} {preprint
  arXiv:1102.4042}\ } (\bibinfo {year} {2011})}\BibitemShut {NoStop}%
\bibitem [{\citenamefont {Rubbo}\ \emph {et~al.}(2012)\citenamefont {Rubbo},
  \citenamefont {Satija}, \citenamefont {Reinhardt}, \citenamefont
  {Balakrishnan}, \citenamefont {Rey},\ and\ \citenamefont
  {Manmana}}]{rubbo2012quantum}%
  \BibitemOpen
  \bibfield  {author} {\bibinfo {author} {\bibfnamefont {C.~P.}\ \bibnamefont
  {Rubbo}}, \bibinfo {author} {\bibfnamefont {I.~I.}\ \bibnamefont {Satija}},
  \bibinfo {author} {\bibfnamefont {W.~P.}\ \bibnamefont {Reinhardt}}, \bibinfo
  {author} {\bibfnamefont {R.}~\bibnamefont {Balakrishnan}}, \bibinfo {author}
  {\bibfnamefont {A.~M.}\ \bibnamefont {Rey}}, \ and\ \bibinfo {author}
  {\bibfnamefont {S.~R.}\ \bibnamefont {Manmana}},\ }\href {\doibase
  10.1103/PhysRevA.85.053617} {\bibfield  {journal} {\bibinfo  {journal} {Phys.
  Rev. A}\ }\textbf {\bibinfo {volume} {85}},\ \bibinfo {pages} {053617}
  (\bibinfo {year} {2012})}\BibitemShut {NoStop}%
\bibitem [{\citenamefont {White}(1992)}]{PhysRevLett.69.2863}%
  \BibitemOpen
  \bibfield  {author} {\bibinfo {author} {\bibfnamefont {S.~R.}\ \bibnamefont
  {White}},\ }\href {\doibase 10.1103/PhysRevLett.69.2863} {\bibfield
  {journal} {\bibinfo  {journal} {Phys. Rev. Lett.}\ }\textbf {\bibinfo
  {volume} {69}},\ \bibinfo {pages} {2863} (\bibinfo {year}
  {1992})}\BibitemShut {NoStop}%
\bibitem [{\citenamefont {Schollw\"ock}(2011)}]{Ulrich201196}%
  \BibitemOpen
  \bibfield  {author} {\bibinfo {author} {\bibfnamefont {U.}~\bibnamefont
  {Schollw\"ock}},\ }\href {\doibase 10.1016/j.aop.2010.09.012} {\bibfield
  {journal} {\bibinfo  {journal} {Ann. Phys.}\ }\textbf {\bibinfo {volume}
  {326}},\ \bibinfo {pages} {96 } (\bibinfo {year} {2011})}\BibitemShut
  {NoStop}%
\bibitem [{\citenamefont {Vidal}(2003)}]{PhysRevLett.91.147902}%
  \BibitemOpen
  \bibfield  {author} {\bibinfo {author} {\bibfnamefont {G.}~\bibnamefont
  {Vidal}},\ }\href {\doibase 10.1103/PhysRevLett.91.147902} {\bibfield
  {journal} {\bibinfo  {journal} {Phys. Rev. Lett.}\ }\textbf {\bibinfo
  {volume} {91}},\ \bibinfo {pages} {147902} (\bibinfo {year}
  {2003})}\BibitemShut {NoStop}%
\bibitem [{\citenamefont {Daley}\ \emph {et~al.}(2004)\citenamefont {Daley},
  \citenamefont {Kollath}, \citenamefont {Schollw\"ock},\ and\ \citenamefont
  {Vidal}}]{daley2004time}%
  \BibitemOpen
  \bibfield  {author} {\bibinfo {author} {\bibfnamefont {A.~J.}\ \bibnamefont
  {Daley}}, \bibinfo {author} {\bibfnamefont {C.}~\bibnamefont {Kollath}},
  \bibinfo {author} {\bibfnamefont {U.}~\bibnamefont {Schollw\"ock}}, \ and\
  \bibinfo {author} {\bibfnamefont {G.}~\bibnamefont {Vidal}},\ }\href@noop {}
  {\bibfield  {journal} {\bibinfo  {journal} {J. Stat. Mech. Theory Exp.}\
  }\textbf {\bibinfo {volume} {2004}},\ \bibinfo {pages} {P04005} (\bibinfo
  {year} {2004})}\BibitemShut {NoStop}%
\bibitem [{\citenamefont {Gobert}\ \emph {et~al.}(2005)\citenamefont {Gobert},
  \citenamefont {Kollath}, \citenamefont {Schollw\"ock},\ and\ \citenamefont
  {Sch\"utz}}]{PhysRevE.71.036102}%
  \BibitemOpen
  \bibfield  {author} {\bibinfo {author} {\bibfnamefont {D.}~\bibnamefont
  {Gobert}}, \bibinfo {author} {\bibfnamefont {C.}~\bibnamefont {Kollath}},
  \bibinfo {author} {\bibfnamefont {U.}~\bibnamefont {Schollw\"ock}}, \ and\
  \bibinfo {author} {\bibfnamefont {G.}~\bibnamefont {Sch\"utz}},\ }\href
  {\doibase 10.1103/PhysRevE.71.036102} {\bibfield  {journal} {\bibinfo
  {journal} {Phys. Rev. E}\ }\textbf {\bibinfo {volume} {71}},\ \bibinfo
  {pages} {036102} (\bibinfo {year} {2005})}\BibitemShut {NoStop}%
\bibitem [{\citenamefont {Manmana}\ \emph {et~al.}(2009)\citenamefont
  {Manmana}, \citenamefont {Wessel}, \citenamefont {Noack},\ and\ \citenamefont
  {Muramatsu}}]{PhysRevB.79.155104}%
  \BibitemOpen
  \bibfield  {author} {\bibinfo {author} {\bibfnamefont {S.~R.}\ \bibnamefont
  {Manmana}}, \bibinfo {author} {\bibfnamefont {S.}~\bibnamefont {Wessel}},
  \bibinfo {author} {\bibfnamefont {R.~M.}\ \bibnamefont {Noack}}, \ and\
  \bibinfo {author} {\bibfnamefont {A.}~\bibnamefont {Muramatsu}},\ }\href
  {\doibase 10.1103/PhysRevB.79.155104} {\bibfield  {journal} {\bibinfo
  {journal} {Phys. Rev. B}\ }\textbf {\bibinfo {volume} {79}},\ \bibinfo
  {pages} {155104} (\bibinfo {year} {2009})}\BibitemShut {NoStop}%
\bibitem [{\citenamefont {Heidrich-Meisner}\ \emph {et~al.}(2009)\citenamefont
  {Heidrich-Meisner}, \citenamefont {Manmana}, \citenamefont {Rigol},
  \citenamefont {Muramatsu}, \citenamefont {Feiguin},\ and\ \citenamefont
  {Dagotto}}]{PhysRevA.80.041603}%
  \BibitemOpen
  \bibfield  {author} {\bibinfo {author} {\bibfnamefont {F.}~\bibnamefont
  {Heidrich-Meisner}}, \bibinfo {author} {\bibfnamefont {S.~R.}\ \bibnamefont
  {Manmana}}, \bibinfo {author} {\bibfnamefont {M.}~\bibnamefont {Rigol}},
  \bibinfo {author} {\bibfnamefont {A.}~\bibnamefont {Muramatsu}}, \bibinfo
  {author} {\bibfnamefont {A.~E.}\ \bibnamefont {Feiguin}}, \ and\ \bibinfo
  {author} {\bibfnamefont {E.}~\bibnamefont {Dagotto}},\ }\href {\doibase
  10.1103/PhysRevA.80.041603} {\bibfield  {journal} {\bibinfo  {journal} {Phys.
  Rev. A}\ }\textbf {\bibinfo {volume} {80}},\ \bibinfo {pages} {041603}
  (\bibinfo {year} {2009})}\BibitemShut {NoStop}%
\bibitem [{\citenamefont {Bethe}(1931)}]{bethe1931theorie}%
  \BibitemOpen
  \bibfield  {author} {\bibinfo {author} {\bibfnamefont {H.}~\bibnamefont
  {Bethe}},\ }\href {\doibase 10.1007/BF01341708} {\bibfield  {journal}
  {\bibinfo  {journal} {Z. Phys. A}\ }\textbf {\bibinfo {volume} {71}},\
  \bibinfo {pages} {205} (\bibinfo {year} {1931})}\BibitemShut {NoStop}%
\bibitem [{\citenamefont {Ovchinnikov}(1967)}]{ovchinnikov1967complexes}%
  \BibitemOpen
  \bibfield  {author} {\bibinfo {author} {\bibfnamefont {A.~A.}\ \bibnamefont
  {Ovchinnikov}},\ }\href
  {http://www.jetpletters.ac.ru/ps/809/article_12463.shtml} {\bibfield
  {journal} {\bibinfo  {journal} {Pis'ma Zh. Eksp. Teor. Fiz.}\ }\textbf
  {\bibinfo {volume} {5}},\ \bibinfo {pages} {48} (\bibinfo {year}
  {1967})}\BibitemShut {NoStop}%
\bibitem [{\citenamefont {Kosevich}\ \emph {et~al.}(1990)\citenamefont
  {Kosevich}, \citenamefont {Ivanov},\ and\ \citenamefont
  {Kovalev}}]{kosevich1990magnetic}%
  \BibitemOpen
  \bibfield  {author} {\bibinfo {author} {\bibfnamefont {A.~M.}\ \bibnamefont
  {Kosevich}}, \bibinfo {author} {\bibfnamefont {B.~A.}\ \bibnamefont
  {Ivanov}}, \ and\ \bibinfo {author} {\bibfnamefont {A.~S.}\ \bibnamefont
  {Kovalev}},\ }\href@noop {} {\bibfield  {journal} {\bibinfo  {journal} {Phys.
  Rep.}\ }\textbf {\bibinfo {volume} {194}},\ \bibinfo {pages} {117} (\bibinfo
  {year} {1990})}\BibitemShut {NoStop}%
\bibitem [{\citenamefont {Matsubara}\ and\ \citenamefont
  {Matsuda}(1956)}]{matsubara1956lattice}%
  \BibitemOpen
  \bibfield  {author} {\bibinfo {author} {\bibfnamefont {T.}~\bibnamefont
  {Matsubara}}\ and\ \bibinfo {author} {\bibfnamefont {H.}~\bibnamefont
  {Matsuda}},\ }\href@noop {} {\bibfield  {journal} {\bibinfo  {journal} {Prog.
  Theor. Phys.}\ }\textbf {\bibinfo {volume} {16}},\ \bibinfo {pages} {569}
  (\bibinfo {year} {1956})}\BibitemShut {NoStop}%
\bibitem [{\citenamefont {Landau}\ and\ \citenamefont
  {Lifschitz}(1935)}]{landau1935theory}%
  \BibitemOpen
  \bibfield  {author} {\bibinfo {author} {\bibfnamefont {L.~D.}\ \bibnamefont
  {Landau}}\ and\ \bibinfo {author} {\bibfnamefont {E.}~\bibnamefont
  {Lifschitz}},\ }\href@noop {} {\bibfield  {journal} {\bibinfo  {journal}
  {Phys. Z. Sowjetunion}\ }\textbf {\bibinfo {volume} {8}},\ \bibinfo {pages}
  {153} (\bibinfo {year} {1935})}\BibitemShut {NoStop}%
\bibitem [{Note3()}]{Note3}%
  \BibitemOpen
  \bibinfo {note} {That is, identifying classical magnetization and momentum
  with their quantum mechanical counter-part.}\BibitemShut {Stop}%
\bibitem [{Note4()}]{Note4}%
  \BibitemOpen
  \bibinfo {note} {It should be noted that the time evolution in Figs.~\ref
  {fig:flip_three}, \ref {fig:inter}(b) and \ref {fig:inter}(c) could have also
  been done using ED instead of t-DMRG.}\BibitemShut {Stop}%
\bibitem [{\citenamefont {Suzuki}(1976)}]{suzuki1976generalized}%
  \BibitemOpen
  \bibfield  {author} {\bibinfo {author} {\bibfnamefont {M.}~\bibnamefont
  {Suzuki}},\ }\href@noop {} {\bibfield  {journal} {\bibinfo  {journal}
  {Commun. Math. Phys.}\ }\textbf {\bibinfo {volume} {51}},\ \bibinfo {pages}
  {183} (\bibinfo {year} {1976})}\BibitemShut {NoStop}%
\bibitem [{\citenamefont {Carr}\ \emph {et~al.}(2001)\citenamefont {Carr},
  \citenamefont {Brand}, \citenamefont {Burger},\ and\ \citenamefont
  {Sanpera}}]{PhysRevA.63.051601}%
  \BibitemOpen
  \bibfield  {author} {\bibinfo {author} {\bibfnamefont {L.~D.}\ \bibnamefont
  {Carr}}, \bibinfo {author} {\bibfnamefont {J.}~\bibnamefont {Brand}},
  \bibinfo {author} {\bibfnamefont {S.}~\bibnamefont {Burger}}, \ and\ \bibinfo
  {author} {\bibfnamefont {A.}~\bibnamefont {Sanpera}},\ }\href {\doibase
  10.1103/PhysRevA.63.051601} {\bibfield  {journal} {\bibinfo  {journal} {Phys.
  Rev. A}\ }\textbf {\bibinfo {volume} {63}},\ \bibinfo {pages} {051601}
  (\bibinfo {year} {2001})}\BibitemShut {NoStop}%
\bibitem [{Note5()}]{Note5}%
  \BibitemOpen
  \bibinfo {note} {It should be mentioned that these wave packets can be
  created directly by a superposition of the specific momentum eigenstate, if
  these are known analytically.}\BibitemShut {Stop}%
\bibitem [{Note6()}]{Note6}%
  \BibitemOpen
  \bibinfo {note} {In order to create more than one localized wave packet, it
  is necessary to remove the interactions (spin hopping) between the parts of
  the chain, where these excitations should be placed. Otherwise all the
  magnetization will fall into one of the valleys of the magnetic field
  potential in the ground state.}\BibitemShut {Stop}%
\bibitem [{\citenamefont {Mikha\u{\i}lov}\ and\ \citenamefont
  {Yaremchuk}(1984)}]{mikhailov1984forced}%
  \BibitemOpen
  \bibfield  {author} {\bibinfo {author} {\bibfnamefont {A.~V.}\ \bibnamefont
  {Mikha\u{\i}lov}}\ and\ \bibinfo {author} {\bibfnamefont {A.~I.}\
  \bibnamefont {Yaremchuk}},\ }\href
  {http://www.jetpletters.ac.ru/ps/88/article_1550.shtml} {\bibfield  {journal}
  {\bibinfo  {journal} {Pis'ma Zh. Eksp. Teor. Fiz.}\ }\textbf {\bibinfo
  {volume} {39}},\ \bibinfo {pages} {296} (\bibinfo {year} {1984})}\BibitemShut
  {NoStop}%
\bibitem [{\citenamefont {Korepin}\ \emph {et~al.}(1997)\citenamefont
  {Korepin}, \citenamefont {Bogoliubov},\ and\ \citenamefont
  {Izergin}}]{korepin1997quantum}%
  \BibitemOpen
  \bibfield  {author} {\bibinfo {author} {\bibfnamefont {V.~E.}\ \bibnamefont
  {Korepin}}, \bibinfo {author} {\bibfnamefont {N.~M.}\ \bibnamefont
  {Bogoliubov}}, \ and\ \bibinfo {author} {\bibfnamefont {A.~G.}\ \bibnamefont
  {Izergin}},\ }\href@noop {} {\emph {\bibinfo {title} {Quantum inverse
  scattering method and correlation functions}}}\ (\bibinfo  {publisher}
  {Cambridge University Press},\ \bibinfo {year} {1997})\BibitemShut {NoStop}%
\bibitem [{\citenamefont {Derzhko}\ \emph {et~al.}(2007)\citenamefont
  {Derzhko}, \citenamefont {Richter}, \citenamefont {Honecker},\ and\
  \citenamefont {Schmidt}}]{derzhko:745}%
  \BibitemOpen
  \bibfield  {author} {\bibinfo {author} {\bibfnamefont {O.}~\bibnamefont
  {Derzhko}}, \bibinfo {author} {\bibfnamefont {J.}~\bibnamefont {Richter}},
  \bibinfo {author} {\bibfnamefont {A.}~\bibnamefont {Honecker}}, \ and\
  \bibinfo {author} {\bibfnamefont {H.-J.}\ \bibnamefont {Schmidt}},\ }\href
  {\doibase 10.1063/1.2780166} {\bibfield  {journal} {\bibinfo  {journal} {Low
  Temp. Phys.}\ }\textbf {\bibinfo {volume} {33}},\ \bibinfo {pages} {745}
  (\bibinfo {year} {2007})}\BibitemShut {NoStop}%
\bibitem [{\citenamefont {Huber}\ and\ \citenamefont
  {Altman}(2010)}]{PhysRevB.82.184502}%
  \BibitemOpen
  \bibfield  {author} {\bibinfo {author} {\bibfnamefont {S.~D.}\ \bibnamefont
  {Huber}}\ and\ \bibinfo {author} {\bibfnamefont {E.}~\bibnamefont {Altman}},\
  }\href {\doibase 10.1103/PhysRevB.82.184502} {\bibfield  {journal} {\bibinfo
  {journal} {Phys. Rev. B}\ }\textbf {\bibinfo {volume} {82}},\ \bibinfo
  {pages} {184502} (\bibinfo {year} {2010})}\BibitemShut {NoStop}%
\end{thebibliography}%

\end{document}